# Exploring Gender and Racial/Ethnic Bias Against Video Game Streamers: Comparing Perceived Gameplay Skill and Viewer Engagement

David Van Nguyen

University of California, Irvine and Microsoft Research, dvnguye5@uci.edu

Edward F. Melcer

University of California, Santa Cruz, eddie.melcer@ucsc.edu

Deanne Adams

Microsoft, dadam@microsoft.com

Research suggests there is a perception that females and underrepresented racial/ethnic minorities have worse gameplay skills and produce less engaging video game streaming content. This bias might impact streamers' audience size, viewers' financial patronage of a streamer, streamers' sponsorship offers, etc. However, few studies on this topic use experimental methods. To fill this gap, we conducted a between-subjects survey experiment to examine if viewers are biased against video game streamers based on the streamer's gender or race/ethnicity. 200 survey participants rated the gameplay skill and viewer engagement of an identical gameplay recording. The only change between experimental conditions was the streamer's name who purportedly created the recording. The Dunnett's test found no statistically significant differences in viewer engagement ratings when comparing White male streamers to either White female ($p = 0.37$), Latino male ($p = 0.66$), or Asian male ($p = 0.09$) streamers. Similarly, there were no statistically significant differences in gameplay skill ratings when comparing White male streamers to either White female ($p = 0.10$), Latino male ($p = 1.00$), or Asian male ($p = 0.59$) streamers. Potential contributors to statistically non-significant results and counter-intuitive results (i.e., White females received non-significantly higher ratings than White males) are discussed.

# 1 INTRODUCTION

Among other marginalized groups, the video games industry has longstanding issues with equity and inclusion for females[1] and underrepresented racial/ethnic minorities. For example, studies suggest that gamers[2] face harassment tied to their gender [11, 23, 45, 51] and race/ethnicity [31, 68, 74, 81]. Moreover, video game characters lack diversity [16, 88, 90]. When female [21, 54, 67] or racial/ethnic minority [7, 35, 50] characters are actually included in games, their portrayals are often flawed. Furthermore, females and certain racial/ethnic minorities are underrepresented among game developers [2, 44], professional gamers [3, 14, 60, 82], and video game streamers [8, 25].

Research suggests there is a perception that females and underrepresented racial/ethnic minorities have worse gameplay skills [22, 77] and produce less engaging video game streaming content [32, 71]. These stereotypes have consequences for diversity, equity, and inclusion in gaming. For example, Freeman and Wohn's [24] study suggests that perceived gameplay skill is an essential selection criteria when recruiting players for eSports teams. Furthermore, experiments suggest that experiencing *gender stereotype threat* may cause female gamers to perform worse in games than they would otherwise [41, 57, 87].

Regarding streaming, these stereotypes may potentially deter females and underrepresented minorities from engaging in streaming [93]. For those who do stream, stereotypes of subpar streaming skill might negatively impact the streamer's audience size [8, 9, 92] and income (e.g., viewers' financial patronage of a streamer and streamer sponsorship offers) [38, 89, 91].

For this paper, we conducted a survey experiment to examine if viewers are biased against video game streamers based on the streamer's gender or race/ethnicity. We specifically examined if there were differences in ratings of the streamer's gameplay skill and viewer engagement. For context, we use Lin et al.'s [47] broad definition of video game streaming: "the practice of (a) publicly performing one's video game play … with (b) varying levels of interactions [which even includes no interaction] between the streamer and spectator and (c) various levels of synchrony" which even includes "Asynchronous [pre]recordings … of game play."

# 2 RELATED WORK

## 2.1 Gender and Racial/Ethnic Bias in Perceived Gameplay Skill

While studies found that males on average have higher gameplay skill than females, this skill gap is negligible or negated when controlling for the confounding variable associated with gender: amount of time that players have devoted to the game [64, 65, 76]. In other words, these studies suggest that (a) the skill gap may be due to females accruing less gameplay time than males and (b) females are not *inherently* less skilled than males.

Even so, multiple qualitative studies suggest that there is a common stereotype within the gaming community that females have less gameplay skill then males [48, 69, 72, 77]. In those studies, gamers proposed two possible causes of the gender skill gap: (a) physiological differences between males and females such as males having faster reaction times, higher spatial intelligence, and higher aggression and (b) societal norms that discourages females and encourages males from gaming, which leads to males accruing more gaming experience [48, 69, 72, 77].

---

[1] Most of the literature on gender bias in gaming seems to focus on cisgender people, which may be distinct from the bias that transgender people face [70].
[2] *Gamer* can be a fraught term. We use a broad definition of the term *gamer*: "people who play video games," which is distinct from whether people *self-identify* with the gamer identity [75].

Prior experiments have directly or indirectly examined gender bias in perceptions of gameplay skill. Kelly et al.'s [42] experiment found that gameplay recordings with female player utterances were rated as less competent than recordings with male utterances. On the other hand, Poeller et al.'s [61] experiment found a main effect that video game streams with female voiceovers received higher ratings than male voiceovers in these gameplay-skill-related dimensions: confidence and competitive. However, there was not a statistically significant main effect on the other gameplay-skill-related dimensions: competence and experience. In Eden et al.'s [18] experiments, research participants were shown gameplay recordings of skilled and unskilled players but participants did not know if the anonymous player was male or female. Research participants then guessed the gender of the player. Eden et al. did *not* find player "skill to be a [statistically] significant predictor of perceived player masculinity." In other words, there was insufficient evidence to support their hypothesis that skilled players are more likely to be perceived as male than unskilled players. On the other hand, Kaye et al. [40] found female players who played as male characters received higher perceived competence ratings than female players who played as female characters. However, there was no statistically significant difference in perceived competence between male players who play as male characters or female characters.

With respect to bias in gameplay skill perceptions of racial minorities, there is relatively limited research. Zhu [94] argues that professional Asian eSport players are perceived as "an immovable, technically skilled behemoth that ultimately dominates the landscape of professional gaming," even surpassing White players. On the other hand, Fletcher [22] argues that there is a meritocracy myth that African American gamers are underrepresented in professional eSports solely because they lack gameplay skills. In other words, Fletcher [22] *seems* to allude to a perception that African American gamers are less skilled than White or Asian gamers.

## 2.2 Gender and Racial/Ethnic Bias in Streamer-Viewer Interactions

Quantitative studies suggest that female streamers are more likely to receive positive chat comments than males [62, 73]. Furthermore, Ruvalcaba et al. [73] did not find a statistically significant difference in negative chat comments between female and male streamers.

On the other hand, qualitative studies suggest that female streamers face gender-based harassment [59, 84, 85]. In fact, while male streamers received receive more game-related chat comments [55], two quantitative studies suggest that female streamers may be more likely to receive objectifying/sexual chat comments [55, 73]. Following up on that last point, one challenge that female streamers face is some viewers' expectations and desire for female streamers to dress revealingly [25]. Female streamers report that some viewers get upset when these inappropriate gendered expectations are not met [25].

Similar to gender, studies also suggest that underrepresented streamers of color experience race-based harassment [84]. For example, one type of harassment is the intentional deployment of emojis that have racist connotations [19]. Furthermore, Han et al.'s [33] research found that streamers who included a stream tag that self-identified themselves as Black or African American were disproportionately targeted in hate raids. Hate raids consist of a coordinated group of human users and/or bots who collectively inundate a streamer's chat with negative comments [33, 52]. Furthermore, Han et al. [33] found that 98% of hate raids in their sample included racist anti-Black chat comments regardless of the streamer's racial/ethnic identity.

## 2.3 Gender and Racial/Ethnic Bias in Perceived Viewer Engagement of Streamers

There are varying conceptualizations of viewer engagement within the literature. Our experiment conceptualizes viewer engagement as distinct from and broader than streamer-viewer interactions (e.g., chat comments): "The concept of [viewer] engagement generally encompasses both cognitive and affective processes and is widely associated with attention, arousal, information interaction, the flow state, aesthetics, novelty, and challenge" [53].

Tangentially related to the concept of viewer engagement, Poeller et al.'s [61] experiment compared personality rating between male and female streamers. However, Poeller et al. did not find a statistically significant main effect on any of these personality-related dimensions: likeable, friendly, or annoying.

Ruberg et al.'s [71] qualitative analysis of forum comments found that female streamers who are perceived to dress revealingly are derisively called *titty streamers* who undeservingly obtain streaming success through sexualized gaming content. In other words, the forum commenters *seem* to perceive that titty streamers produce less engaging content than males and thus have to resort to sexualizing themselves. Furthermore, commenters argued that titty streamers delegitimize the work of non-sexual female streamers because it (a) shapes viewers' perception that female streamers (including non-sexual females) are generally less talented than males and (b) encourages viewers to sexually harass non-sexual female streamers.

Similarly, Gray's [32] qualitative analysis of forum comments suggest viewers perceive that Black streamers are less engaging than White streamers. Specifically, commenters said that Black streamers (a) are "too urban" and fall outside the norm of the typical White streamer and (b) may not appeal to Twitch viewers who are predominately White. In fact, one quantitative study suggests that there may be demographic homophily in Twitch viewership (i.e., viewers are drawn to streamers who have similar demographic characteristics) [83].

The following papers do not explicitly study viewer engagement of Asian streamers but still provide useful related information. Zhu [94] argues that the North American and European gaming community dehumanizes/robotizes and emasculates Asian eSports players in order to relieve White anxiety over Asian dominance in eSports competitions. Perhaps, this framing of Asians may result in lower viewer engagement of Asian eSport players. On the other hand, Fickle et al. [20] argues that "games themselves have become Asiatic products even when they contain no explicit racial representations, as they are manufactured and innovated in Asian contexts and are often concerned with, and played by, sizable Asian audiences." So, perhaps viewers perceive Asian streamers to be more engaging because games are associated with Asia.

## 2.4 Gaps in the Literature

As shown in the related work, most studies on bias in perceived gameplay skill and viewer engagement use non-experimental methods. An exception is the four experimental papers that examined gender bias in perceived gameplay skill [18, 40, 42, 61]. So, our experiment is not novel in that regard. However, to our knowledge, our experiment is the first experiment to (a) examine racial/ethnic bias in perceived gameplay skill or (b) examine gender or racial/ethnic bias in perceived viewer engagement of streamers.

## 3 METHODOLOGY

Our between-subjects survey experiment examined if viewers had gender or racial/ethnic bias against video game streamers. Survey participants watched an identical gameplay recording. The only aspect that changed between experimental conditions was the streamer's name who purportedly created the gameplay recording, which was either a White male name, White female name, Asian male name, or Latino male name.[3] Based on the recording, survey participants then rated the streamer's gameplay skill and viewer engagement.

### 3.1 Participants

During the spring 2023, summer 2023, and fall 2023 term, we emailed University of California, Irvine (UCI) and University of California, Santa Cruz (UCSC) instructors and asked them to forward the call for study participants to their students. During each academic term, we compiled a list of instructors from the university class schedule. In total, we sent an initial email and a follow up email to 1,059 unique UCI instructors and 813 unique UCSC instructors.[4]

The survey participant eligibility requirements include: (a) be a UCI or UCSC student and (b) have experience playing *The Legend of Zelda: Breath of the Wild*. To mitigate bots, participants were required to login using a valid UCI or UCSC single sign-on username and password to access the Qualtrics survey. From there, survey participants who did not meet participant eligibility requirements or did not provide their research consent were redirected out of the survey.

We preplanned to collect 200 *unexcluded* responses. We received 215 completed survey responses. 15 responses were excluded based on the exclusion criteria in Appendix C. The 200 unexcluded participants' demographics are provided in Table 1. We did not provide participant compensation because the survey is relatively short and compensation can incentivize fraudulent responses from humans or bots [80].

---

[3] While most of the literature on racial bias in video games focuses on Black gamers, we chose not to include an African American male experimental condition. Instead, we used the Latino male experimental condition to represent bias against underrepresented racial/ethnic minority males. Within the US context, common African American last names like Washington, Jefferson, and Jackson are racially ambiguous [26, 30]. For example, Gaddis [26] found that White first names paired with common African American last names (e.g., Brian Washington) *were perceived as White 67.9%* of the time by study participants. Another option is to use racially distinctive African American first names such as DaShawn and Jamal [26] but that may introduce confounds such as socioeconomic status [12].

[4] If a professor was already contacted during a prior term, then they were excluded from any additional research outreach in future terms.

Table 1: Participant Demographics within Each Experimental Condition

|  | White Male Streamer | White Female Streamer | Latino Male Streamer | Asian Male Streamer |
|---|---|---|---|---|
| Gender |  |  |  |  |
|   Female | 29% | 21% | 30% | 21% |
|   Male | 66% | 71% | 63% | 71% |
|   Non-binary / third gender | 3% | 5% | 5% | 5% |
|   Prefer to self-describe | 3% | 3% | 3% | 2% |
| Race and ethnicity |  |  |  |  |
|   American Indian or Alaska Native | 1% | 0% | 3% | 0% |
|   Asian | 33% | 39% | 55% | 36% |
|   Black or African American | 4% | 5% | 0% | 7% |
|   Hispanic or Latino | 26% | 24% | 18% | 24% |
|   Middle Eastern or North African | 5% | 3% | 0% | 0% |
|   Pacific Islander | 3% | 0% | 0% | 5% |
|   White | 56% | 50% | 48% | 50% |
|   Other | 0% | 5% | 0% | 0% |
| Age | 20.1 ± 2.8 | 23.1 ± 13.1 [a] | 19.4 ± 1.7 | 19.9 ± 2.0 |
| Experience Playing *Breath of the Wild* | 4.0 ± 0.8 | 3.9 ± 0.9 | 4.2 ± 0.9 | 3.8 ± 0.8 |
| Experience Watching Gameplay Videos [b] | 4.3 ± 0.9 | 4.3 ± 0.9 | 4.4 ± 0.7 | 4.3 ± 0.9 |

*Note*. The last three rows display means and standard deviations. The two experience-related rows use 5-point unipolar response options ranging from "Not experienced at all" (1) to "Extremely experienced" (5). [a] The White female experimental condition's age standard deviation is relatively large because there are several nontraditional age survey participants. [b] Participants' experience with watching gameplay videos includes gameplay videos for any game (not just *Breath of the Wild*).

### 3.2 Materials

Before deploying the survey experiment, we ran one-on-one 30-minute pilot test sessions with six gamers who played *Breath of the Wild*. We then revised the experiment based on their feedback.

#### 3.2.1 Streamer Names

Our experiment used names to signal the gender and race/ethnicity of the streamer. Crabtree et al. [13] recommend that experiments that use names should employ stimulus sampling. Stimulus sampling is "a procedure for increasing the generalizability of research results by using multiple stimuli within a category as representative of an experimental condition, as opposed to selecting a single stimulus whose unique characteristics may distort results" [1]. Accordingly, after survey participants were randomly assigned to one of the four experimental conditions, they were then randomly assigned to one of four names that are representative of that experimental condition (see Table 2). For example, once a survey participant was randomly assigned to the Latino male streamer experimental condition, they were then randomly assigned one of these four representative Latino male names: Brian Velazquez, Daniel Orozco, Robert Hernandez, or Steven Gonzalez.

Table 2: Names Used to Signal the Video Game Streamer's Gender and Race/Ethnicity

| White Male Names | White Female Names | Latino Male Names | Asian Male Names [a] |
| --- | --- | --- | --- |
| Brian Nielsen | Amy Nielsen | Brian Velazquez | Albert Chen |
| Daniel McGrath | Erica McGrath | Daniel Orozco | Andrew Wang |
| Robert Becker | Laurie Becker | Robert Hernandez | Eric Yang |
| Steven Andersen | Stephanie Andersen | Steven Gonzalez | Peter Li |

*Note.* After survey participants were randomly assigned to one of the four experimental conditions, they were then randomly assigned to one of four names that are representative of that experimental condition. [a] The Asian names [13] come from a different dataset of validated names than the White and Latino names [26, 27].

We used full names from validated datasets that measured names' race/ethnicity signaling [13, 26, 27]. (Refer to Appendix A for details on the racial/ethnic perceptions and limitations of these validated full names.) Across all experimental conditions, we chose full names that were comprised of a typically White first name paired with either a White last name, Latino last name, or an Asian last name (see Table 2). In other words, instead of using a Latino first name with a Latino last name (e.g., Javier Velazquez), we paired a typically White first name with a Latino last name (e.g., Brian Velazquez) in an attempt to control for confounds such as US generation status and assimilation [28, 49].

Study participants are exposed to the streamer's name in the instructions before the gameplay recording and within each survey rating question (e.g., "Rate [STREAMER_NAME]'s skill in archery."). However, the streamer's name was *not* embedded within the gameplay recording video.

### 3.2.2 Gameplay Recording

We chose *The Legend of Zelda: Breath of the Wild*, a single-player action-adventure game, for our experiment's gameplay recording. We chose this game because (a) it is popular and therefore may have a larger pool of study participants to draw from and (b) the computer-controlled enemies have a consistent attack pattern, which may make it easier for viewers to assess the player's gameplay skill. Furthermore, the game provides obvious visual and audio cues when players successfully execute advanced combat actions, which require precise timing and knowledge of computer-controlled enemy attack patterns. For example, when players execute a "Perfect Dodge" there is a slow-motion visual effect and a whoosh sound effect. (Refer to the following footnotes for our rationale for not choosing a multi-player game[5] or not choosing a game with objective in-game performance metrics[6] for our experiment.)

Our experiment's *Breath of the Wild* gameplay recording was sourced from a pre-existing public YouTube video (i.e., the paper authors did *not* create the video) [10]. The asynchronous gameplay recording does not include a camera view of a streamer or a streamer voiceover (i.e., the streamer did not use their webcam or microphone). Furthermore, there is no streamer-audience interaction within the pre-recorded video. (Even with all these restrictions, this pre-recorded YouTube video still fits within Lin et al.'s [47] broad definition of game streaming.)

---

[5] Gameplay skill is often studied within the context of competitive multi-player games [15, 37, 46]. However, we purposely avoided choosing a multiplayer game because it may be difficult for non-professional viewers to assess gameplay skill when there are multiple human players in a decontextualized 1-minute video recording [66].

[6] We avoided games with objective in-game performance metrics as it may influence viewers' perceptions (e.g., reduce bias in viewers' subjective ratings). For example, the game *Hi-Fi Rush* uses letter grade performance metrics. If a *Hi-Fi Rush* gameplay video had an in-game letter grade of "C," then survey participants may feel less justified to give a White male streamer a very high subjective rating or a Latino male streamer a very low subjective rating. Then, statistically significant differences would be harder to detect.

Within the 1-minute recording, the streamer defeats a Lynel, which is generally recognized as a difficult enemy in *Breath of the Wild*. However, the streamer gets damaged by the Lynel once and the streamer does not execute any advanced combat actions (e.g., "Perfect Guard," "Perfect Dodge," "Flurry Rush," etc.).

Survey participants were told that the gameplay recording was a YouTube video and that the streamer was a YouTuber. We intended for the YouTube video to be perceived as having medium gameplay skill and medium level of viewer engagement. This seems to align with our survey participants' perceptions. When aggregating all survey participants together ($N = 200$), the gameplay recording received an average rating of 2.51 out of 5 ($SD = 0.71$) in the viewer engagement survey scale and an average rating of 2.78 out of 5 ($SD = 0.58$) in the gameplay skill survey scale (see Appendix B).

*3.2.3 Survey*

The Qualtrics survey contained 16 questions and consisted of the following parts in this order: consent form and one research consent question, one study eligibility question, an embedded *Breath of The Wild* gameplay recording, the first attention check question, four gameplay skill rating questions, the second attention check question, four viewer engagement rating questions, and four survey participant demographic questions.

Gameplay skill is comprised of multiple factors, which can vary in importance based on the game genre [58]. However, the *Breath of the Wild* gameplay recording (and consequently the gameplay skill survey scale) mostly focuses on one factor, which Norman et al. [58] call perceptual-motor abilities. Our gameplay skill survey scale asked participants to rate the YouTuber in terms of (a) cheapness, (b) dodging and guarding skills, (c) archery skills, and (d) melee attack skills.

Our viewer engagement survey scale was influenced in part by Hilvert-Bruce et al.'s [36] study of viewers' motivations behind Twitch *live stream* engagement. For our *asynchronous* pre-recorded YouTube video – which did not include streamer-viewer interactions – the most relevant motivations from Hilvert-Bruce et al.'s [36] study were entertainment and information seeking. Our viewer engagement survey scale asked participants to rate the gameplay recording based on the degree of (a) entertainment, (b) informativeness, (c) like or dislike, and (d) recommendation of the video.

See Appendix B for the survey scale items and response options within the gameplay skill scale and viewer engagement scale.

**3.3 Analysis**

Our experiment was preregistered at AsPredicted (https://aspredicted.org/y2c5x.pdf). We followed all aspects of our preregistration except for one survey exclusion criteria (see Appendix C).

Statistical tests were run on SPSS Statistics (Version 29). To measure the internal consistency of our multi-item survey scales, we preregistered that an acceptable Cronbach's alpha value is 0.7 or higher. For comparing means, we preregistered using one-way ANOVA and two-tailed Dunnett's test with a significance level of 0.05.

As opposed to all pairwise comparisons, we chose Dunnett's test to increase statistical power. The Dunnett's test compares the control group to all the other experimental groups. We designated the White male streamer as the control group. To increase statistical power, we used the square root allocation rule [56] for Dunnett's test to randomly allocate approximately 40% of study participants to the control group (i.e., White male streamer) and approximately 20% to each of the three other experimental groups.

# 4 RESULTS

Our viewer engagement survey scale has an acceptable internal consistency ($\alpha = 0.78$, 4 questions). For the one-way ANOVA, there was sufficient evidence to support the alternative hypothesis that the group means for viewer engagement ratings are not all equal: $F(3, 196) = 3.19, p = 0.02, \eta^2 = 0.05$. However, the Dunnett's test (see Table 3) found no statistically significant differences in viewer engagement ratings when comparing White male streamers to either White female ($p = 0.37$), Latino male ($p = 0.66$), or Asian male ($p = 0.09$) streamers.

Table 3: Dunnett's Test for Viewer Engagement Survey Scale

| Comparison | Control | | Treatment | | $p$ | $d$ | 95% CI for $d$ |
|---|---|---|---|---|---|---|---|
| | $M$ | $SD$ | $M$ | $SD$ | | | |
| WM - WF | 2.43 | 0.69 | 2.63 | 0.76 | 0.37 | -0.28 | [-0.66, 0.11] |
| WM - LM | 2.43 | 0.69 | 2.30 | 0.60 | 0.66 | 0.20 | [-0.18, 0.58] |
| WM - AM | 2.43 | 0.69 | 2.72 | 0.73 | 0.09 | -0.41 | [-0.78, -0.03] |

*Note.* White male (WM) streamer is the control group that is being compared to the treatment groups: White female (WF), Latino male (LM), and Asian male (AM) streamer.

Our gameplay skill scale has poor internal consistency ($\alpha = 0.53$, 4 questions). For the one-way ANOVA, there was not sufficient evidence against the null hypothesis that the group means for gameplay skill ratings are all equal: $F(3, 196) = 1.76, p = 0.16, \eta 2 = 0.03$. Similarly, the Dunnett's test (see Table 4) found no statistically significant differences in gameplay skill ratings when comparing White male streamers to either White female ($p = 0.10$), Latino male ($p = 1.00$), or Asian male ($p = 0.59$) streamers.

Table 4: Dunnett's Test for Gameplay Skill Survey Scale

| Comparison | Control | | Treatment | | $p$ | $d$ | 95% CI for $d$ |
|---|---|---|---|---|---|---|---|
| | $M$ | $SD$ | $M$ | $SD$ | | | |
| WM - WF | 2.71 | 0.56 | 2.95 | 0.75 | 0.10 | -0.38 | [-0.77, 0.01] |
| WM - LM | 2.71 | 0.56 | 2.71 | 0.55 | 1.00 | -0.01 | [-0.39, 0.37] |
| WM - AM | 2.71 | 0.56 | 2.83 | 0.44 | 0.59 | -0.23 | [-0.61, 0.14] |

*Note.* White male (WM) streamer is the control group that is being compared to the treatment groups: White female (WF), Latino male (LM), and Asian male (AM) streamer.

The original gameplay skill scale ($\alpha = 0.53$, 4 questions) asked users to rate the streamer on cheapness, dodging skill, archery skill, and melee skills (see Appendix B). However, the cheap rating survey question has low correlation (ranging from 0.05 to 0.12) with the other gameplay skill scale questions. After removing the cheap rating question, the *modified* gameplay skill scale then had an increased Cronbach's alpha of 0.65 with the three remaining survey questions.[7]

We then ran one-way ANOVA and Dunnett's test with the modified gameplay skill scale, which was *not* preplanned in our study preregistration. For the one-way ANOVA, there was not sufficient evidence against the null hypothesis that the group means for the modified gameplay skill ratings are all equal: $F(3, 196) = 2.20$, $p = 0.09$, $\eta2 = 0.03$. Similarly, the Dunnett's test (see Table 5) found no statistically significant differences in modified gameplay skill ratings when comparing White male streamers to either White female ($p = 0.09$), Latino male ($p = 1.00$), or Asian male ($p = 0.34$) streamers.

---

[7] Some researchers consider an acceptable Cronbach's alpha value to be 0.65 or higher [86]. This lower threshold value may be acceptable considering that the modified gameplay skill scale only has three survey questions. All else held equal, Cronbach's alpha increases as the number of scale survey questions increases [86]. However, to be clear, we preregistered an acceptable Cronbach's alpha value to be 0.7 or higher.

Table 5: Dunnett's Test for the Modified Gameplay Skill Survey Scale

| Comparison | Control | | Treatment | | $p$ | $d$ | 95% CI for $d$ |
|---|---|---|---|---|---|---|---|
| | $M$ | $SD$ | $M$ | $SD$ | | | |
| WM - WF | 2.28 | 0.63 | 2.55 | 0.87 | 0.09 | -0.39 | [-0.78, 0.00] |
| WM - LM | 2.28 | 0.63 | 2.26 | 0.60 | 1.00 | 0.03 | [-0.35, 0.41] |
| WM - AM | 2.28 | 0.63 | 2.46 | 0.50 | 0.34 | -0.31 | [-0.69, 0.06] |

*Note.* The modified gameplay skill survey scale removed the cheap rating survey question. White male (WM) streamer is the control group that is being compared to the treatment groups: White female (WF), Latino male (LM), and Asian male (AM) streamer.

## 5 DISCUSSION

To summarize, our Dunnett's tests found no statistically significant differences in gameplay skill ratings or viewer engagement ratings when comparing White male streamers to either White female, Latino male, or Asian male streamers. In other words, there was not sufficient evidence against the null hypotheses that White male streamers receive equal mean ratings compared to either White female, Latino male, or Asian male streamers.

### 5.1 Potential Contributors to Counterintuitive Gender-Specific Results

We initially hypothesized that White male streamers would receive higher ratings than White females. However, although not statistically significant, White female streamers counterintuitively received higher ratings on viewer engagement ($p = 0.37$, $d = -0.28$, 95% CI [-0.66, 0.11]) and gameplay skill ($p = 0.10$, $d = -0.38$, 95% CI [-0.77, 0.01]) compared to White males.

Our gender-specific results diverge from (a) experiments that found male players and male avatars were perceived to have higher gameplay competency than their female counterparts [40, 42] and (b) a qualitative study that suggests viewers perceive that female streamers produce less engaging streaming content than males [71]. On the other hand, our gender-specific results align with an experiment that found female streamers received higher ratings than male streamers in a subset of gameplay-skill-related dimensions [61].

Several studies provide potential explanations behind our counterintuitive gender-specific results. For example, a few semi-professional and professional gamers in an interview study reported that males are held to a higher expectation/standard for gameplay skill compared to females [48]. Similarly, competitive-level female gamers are often told "you're pretty good for a girl"; that phrase implies lower expectations/standards for females [4]. So, perhaps White male streamers in our study received lower ratings because survey participants have different performance expectations for male versus female players to be considered skilled.

Other studies suggest that male gamers engage in *white knighting*, where males provide unsolicited help/support to females because they (a) want to counteract the gender-based toxicity that female gamers experience or (b) may implicitly subscribe to the infantilizing belief that females are bad at gaming and therefore need help [5, 48]. In other words, perhaps White female streamers in our experiment received higher ratings because study participants wanted to be supportive of female streamers.

## 5.2 Potential Contributors to Statistically Non-Significant Results

Our experiment's non-significant results diverge from the majority of *related* experiments. Four experimental papers have examined gender bias in perceptions of gameplay skill: three papers had statistically significant results [40, 42, 61] and one paper lacked statistically significant results [18]. (To our knowledge, there are no prior published experiments that (a) examine racial/ethnic bias in perceived gameplay skill or (b) examine gender or racial/ethnic bias in perceived viewer engagement of streamers.)

There are various potential contributors to our lack of statistically significant results. First, studies suggest college students tend to be more liberal than non-college-educated people [6, 34, 39]. Perhaps our survey participants, who were California college student gamers, might be more progressive in their views of gender and race/ethnicity than the typical American gamer.

Second, our experiment differed from other video game bias experiments in how we signaled demographic characteristics. Our experiment used first and last names to signal the streamer's gender and race/ethnicity. To signal the gamer's gender, Kaye et al.'s [40] experiment similarly used first names (e.g., "Alice") but also included gender pronouns (e.g., "her"). On the other hand, other experiments used female or male voiceovers (e.g., "oh no," "argh," or "yes") within a gameplay recording to signal the gamer's gender [42, 61]. Perhaps reading a gamer's name is a less evocative trigger for bias than other signals like hearing the gamer's voice or seeing the gamer's appearance. In other words, the gamer's gender and race/ethnicity might be more present in research participants' thought process if they hear or see the gamer as opposed to reading the gamer's name.

Third, we used a Latino male streamer experimental condition instead of an African American male streamer experimental condition due to the racial ambiguity of common African American last names such as Washington, Jefferson, and Jackson [26, 30]. However, a meta-analysis of correspondence audit experiments suggests that Hispanic Americans experience less severe racial/ethnic discrimination compared to African Americans within certain contexts (e.g., rental housing) [29]. Perhaps perceptions of video-game-related skills are another context where Latino Americans experience less severe discrimination compared to African Americans. For example, while not an experiment, Han et al.'s [33] research suggests that (a) Black Twitch streamers are disproportionately targeted in hate raids and (b) hate raids most commonly contain racist anti-Black chat comments regardless of the streamer's racial/ethnic identity. Whereas it seems like Latinos/as were not a main target in the hate raids within Han et al.'s [33] research sample. In other words, we may have seen a larger effect size and achieved statistically significant results if we had used an African American male streamer experimental condition instead of the Latino male streamer experimental condition.

## 5.3 Limitations

There are several noteworthy limitations. First, the experiment's gameplay recording did not include the streamer's webcam or microphone, which may not be generalizable to video game streams that do use those equipment [17, 78]. This was done to avoid the known and unknown confounding variables that come from hiring actors to play the role of White male, White female, Latino male, or Asian male streamer (e.g., despite efforts to recruit actors of similar subjective attractiveness, unknowingly one actor may *still* have a more attractive appearance).

Second, the experiment's *asynchronous* pre-recorded YouTube video – which did not include streamer-viewer interactions – may not be generalizable to *live* streams that feature streamer-viewer interactions [47, 79].

Third, our experiment only used one game (*Breath of the Wild*). As such, our results may not be generalizable to other games.

Fourth, we do not know if survey participants watched the full 1-minute-long video because we only recorded how much time participants spent on the overall survey. In retrospect, we should have also added a Qualtrics timer to record how much time participants spent specifically on the video.

Fifth, we do not know if survey participants paid attention while watching the video. Granted, we did ask survey participants "To verify that you are not a robot, what is the environment in the Breath of the Wild video?" However, a survey participant would just need to watch a few seconds of the video to correctly identify the environment. Instead, we should have asked a question that would force the survey participant to carefully watch the full video in order to answer the attention check question correctly.

Sixth, although the streamer's name was included within each survey rating question, we are unsure if survey participants paid attention to or were even aware of the streamer's name.

Seventh, although we used names from validated name datasets [13, 26, 27], we are unsure if survey participants perceived the names' gender and race/ethnicity as we intended. To avoid raising suspicion about the true intent of the study, we did not incorporate a manipulation check into the survey (i.e., we did not ask survey participants to identify/guess the streamer's gender or race/ethnicity based on their name). Furthermore, study participants were only debriefed about the true intent of the study after all data collection was complete. We made these study design choices because gender and racial/ethnic bias in gaming can be a controversial topic for some gamers (i.e., survey participants who become upset may tip off *prospective* participants about the true intent of the study).

Eighth, we use a different set of *first* names (see Table 2) for the Asian male streamer experimental condition (i.e., Albert, Andrew, Eric, and Peter) compared to the White male and Latino male streamer experimental condition (i.e., Brian, Daniel, Robert, and Steven). This discrepancy is because the Asian names [13] come from a different dataset of validated names than the White and Latino names [26, 27]. There may have been confounds with using differing first names (e.g., perhaps one set of first names are perceived to be of higher socioeconomic status than the other).

### 5.4 Future Research Directions

We suggest several areas for future research on measuring bias in the video game community. First, researchers can study bias of gamers in different settings and modes: (a) instead of measuring gameplay skill within a single-player game, researchers can examine bias using a competitive multi-player game and (b) instead of measuring viewer engagement of an asynchronous pre-recorded YouTube video, researchers can use a recording of a live-stream where the streamer is interacting with their audience. Second, aside from females and racial/ethnic minorities, researchers can study bias against other marginalized groups such as LGBT or older gamers. Third, aside from studying streamers, researchers can run experiments that examine bias against different video game professions such as developers, journalists, and eSports commentators/announcers.

# APPENDICES

**Appendix A: Racial Perceptions and Limitations of Names from Validated Datasets**

We used names (see Table 2) from validated datasets that measured names' race/ethnicity signaling [13, 26, 27]. However, a limitation that is shared by all aforementioned datasets is that they did not measure gender perceptions of names in their dataset.

The White male names, White female names, and Latino male names (see Table 2) used in our experiment were pulled from Gaddis' [26, 27] dataset of validated names. Within his paper, Gaddis did not list *specific* congruent race/ethnicity perception rates on a per-full-name basis (e.g., he did not list the specific congruent rate for Brian Nielsen). He instead listed the *overall* congruent perception rates for name pairing types. The overall set of White first names with White last names were perceived as White 92% of the time by study participants. On the other hand, the overall set of White first names with Latino last names were perceived as Latino 75% of the time.

Gaddis [26, 27] also categorized first names in his dataset by lower, middle, and upper socioeconomic status (SES), which was based on New York birth records of mothers' education level. For our study's White male name, White female name, and Latino male name, we chose first names that were categorized as middle SES. However, to be clear, Gaddis did not measure the SES *perceptions* of the names in his dataset.

The four Asian male names (see Table 2) used in our experiment were pulled from Crabtree et al.'s [13] dataset of validated names. This dataset asked study participants to guess the race/ethnicity, citizenship, income, and highest level of education of full names. Crabtree et al. provided data on a per-full-name basis (e.g., they listed the specific perception data for Albert Chen). On average, these four specific Asian male full names were perceived as Asian 77% of the time and perceived as U.S. citizens 83% of the time. On average, the four Asian male full names were perceived as approximately middle income and having a bachelor's degree as their highest degree.

**Appendix B: Items in Multi-Item Survey Scales**

Table B1: Items in Gameplay Skill Scale

| Scale Item | Response Options [a] |
|---|---|
| Was it cheap for [STREAMER_NAME] to repeatedly use "Stasis" to freeze the Lynel? For the purposes of this study, "cheap" is defined as using overpowered combat actions that are easy to execute. | Not cheap at all (5) – Extremely cheap (1) [b] |
| Rate [STREAMER_NAME]'s skill in dodging and guarding enemy attacks. Rating criteria include how many times the player was damaged, player awareness of the Lynel's predictable attack patterns, and executing advanced techniques (e.g., "Perfect Dodge"). | Not skilled at all (1) – Extremely skilled (5) |
| Rate [STREAMER_NAME]'s skill in archery. Rating criteria include percent of shots that hit any part of the Lynel, percent of shots that specifically are "Critical Hit" headshots, and executing advanced techniques (e.g., shooting arrows while Link is in mid-air). | Not skilled at all (1) – Extremely skilled (5) |
| Rate [STREAMER_NAME]'s skill in melee attacks with a sword. Rating criteria include percent of successful melee attacks, effectiveness of melee attacks, and executing advanced melee techniques (e.g., "Flurry Rush"). | Not skilled at all (1) – Extremely skilled (5) |

*Note*. These survey questions included hyperlinks to wiki pages that described these *Breath of the Wild* combat terminology. [a] The column shows the text and numeric score of the first and last response option. Each scale item uses 5-point unipolar response options. [b] This scale item was reverse scored because cheap is a negative gaming trait.

Table B2: Items in Viewer Engagement Scale

| Scale Item | Response Options [a] |
|---|---|
| How entertaining was [STREAMER_NAME]'s YouTube video? | Not entertaining at all (1) – Extremely entertaining (5) |
| Was [STREAMER_NAME]'s YouTube video a useful demonstration of how to defeat a Lynel? | Not useful at all (1) – Extremely useful (5) |
| Do you like or dislike [STREAMER_NAME]'s YouTube video? | Dislike a great deal (1) – Like a great deal (5) [b] |
| Imagine that your friend is looking for highly skilled Breath of the Wild gameplay videos. How strongly would you recommend [STREAMER_NAME]'s video to your friend? | Not recommended at all (1) – Extremely recommended (5) |

*Note*. [a] The column shows the text and numeric score of the first and last response option. [b] This scale item uses 5-point *bipolar* response options. All other scale items use 5-point *unipolar* response options.

**Appendix C: Survey Exclusion Criteria**

Of the fully completed survey responses, we excluded responses if they met any of the following exclusion criteria:

- Answered the honeypot survey question, which are only viewable and answerable by bots [80].
- Failed any of the two attention check questions: "To verify that you are not a robot, what is the environment in the Breath of the Wild video?" and "To verify that you are not a robot, please select: 'Very challenging.'"
- Typed an impossible human age or typed irrelevant text when choosing the option to self-describe their gender or race/ethnicity.
- Flagged as a human who retook the survey multiple times or flagged as a bot.[8]
- Flagged as a speeder, which we define as responses that are faster than one standard deviation from the median duration. Our speeding cutoff point is more stringent than Qualtrics' definition of speeders: "more than two standard deviations from the median duration" [63].

We also preplanned to exclude survey respondents who engaged in straightlining. However, we did not use this exclusion criteria because our survey scale items do not have uniform response options (see Appendix B). In other words, our multi-item survey scales do not meet the typical definition of straightlining: "Straightlining occurs when survey respondents give identical (or nearly identical) answers to items in a battery of questions using the same response scale" [43].

---

[8] Whereas both the UCI and UCSC Qualtrics license provided embedded reCAPTCHA v3 functionality, only the UCSC Qualtrics license provided embedded RelevantID functionality [63].